# Precise Tool to Target Positioning Widgets (TOTTA) in Spatial Environments: A Systematic Review

Mine Dastan ⓘ, Michele Fiorentino ⓘ and Antonio E. Uva ⓘ

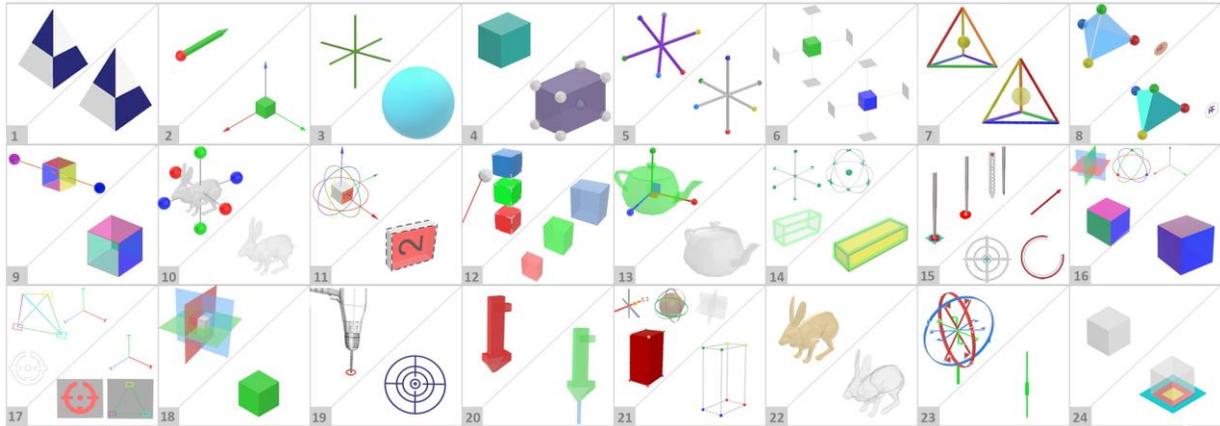

Fig. 1. The Tool to Target (TOTTA) widget designs from the 24 papers analyzed chronologically. For each frame, the Tool widget (TO) is depicted at the top left, and the Target widget (TA) at the bottom right.

**Abstract**—TOTTA outlines the spatial position and rotation guidance of a real/virtual tool (TO) towards a real/virtual target (TA), which is a key task in Mixed reality applications. The task error can have critical consequences regarding safety, performance, and quality, such as surgical implantology or industrial maintenance scenarios. The TOTTA problem lacks a dedicated study and it is scattered in different domains with isolated designs. This work contributes to a systematic review of the TOTTA visual widgets, studying 70 unique designs from 24 papers. TOTTA is commonly guided by the visual overlap –an intuitive, pre-attentive "collimation" feedback– of simple shaped widgets: Box, 3D Axes, 3D Model, 2D Crosshair, Globe, Tetrahedron, Line, Plane. Our research discovers that TO and TA are often represented with the same shape. They are distinguished by topological elements (e.g. edges/vertices/faces), colors, transparency levels, and added. shapes, widget quantity, and size. Meanwhile some designs provide continuous "during manipulation feedback" relative to the distance between TO and TA by text, dynamic color, sonification, and amplified graphical visualization. Some approaches trigger discrete "TA reached feedback" such as color alteration, added sound, TA shape change, and added text. We found the lack of golden standards, including in testing procedures, as current ones are limited to partial sets with different and incomparable setups (different target configurations, avatar, background, etc.). We also found a bias in participants: right-handed, young male, non-color impaired.

✦

## 1 INTRODUCTION

Mixed reality (MR) and relative spatial environments are increasing in popularity thanks to the steady evolution of technology and the reduced costs of head-mounted displays, improved tracking, software ecosystems like SteamVR, and development tools such as Unity3D [1, 2]. However, MR spatial environment management is rather limited to gaming and entertainment. Among the key reasons for the lack of adoption in "serious" applications is the inadequate performance of spatial interactions [3, 4, 5]. Many industrial applications (surgery, maintenance, welding, etc.) require the user to handle a tool and perform spatial dexterity tasks with 6 degrees of freedom (6 DOF, three positional and three rotational), with a low margin of error to avoid consequences in terms of cost, time, and even human safety [6], 7, 8, 9]. MR has been demonstrated to be beneficial in training and guiding spatial operations using virtual reality (VR) and executing in the real world using augmented reality (AR) [10, 11, 12]. These spatial operations are challenging since they demand precision manipulation of virtual/real tools [13, 14, 15].

Spatial XR interactions used in wide application fields require precise tool-to-target (TOTTA) tasks (position/rotation). In medical surgical procedures, where implementing haptic feedback is complex and costly [16, 17, 18, 19] and the surgeon must place the tool accurately on the preoperative target. Moreover, in industrial scenarios, tasks such as quality checks of diagnostic sensors to verify carbon fiber structures [20, 21, 22] requires a certain precision.

Although a 3D tool to target positioning tasks is important, there is not a specific definition of this problem that differentiates it from general 2D/3D UI research like "drag-and-drop," "aiming/aligning," and "docking" [23]. Therefore, we define it as the TOTTA (TO-to-TA), a spatial positioning task. TOTTA is usually guided by visual elements -the widgets [24, 25]- to be aligned by spatial skills in 3D space. These widgets help manipulate the tool (physical or virtual TO) towards a spatial target (physical or virtual TA). The user manipulates a tool towards a target, supported by tool widgets (TO) and target widgets (TA), to achieve higher precision and performance.

- *Mine Dastan, Antonio E. Uva and Michele Fiorentino are with the Polytechnic University of Bari. E-mail: mine.dastan@poliba.it*

Table 1: Summary of the selected papers' used technology, interaction, visualization, and tracking device.

| No. | Ref. | Authors | Technology | Interaction | Visualization Device | Tracking Device | Field |
|---|---|---|---|---|---|---|---|
| 1 | [26] | Boritz et al. | Desktop VR | Object-Mouse | CrystalEyesT | Fastrak | Near |
| 2 | [27] | Fiorentino et al. | Projected VR HMD | Controller | Vertical screen, Projectors | ART | Near |
| 3 | [28] | Veit et al. | VR HMD | Gesture-Touch | CrystalEyes CE-2 | ART | Far |
| 4 | [29] | Ragan et al. | VR HMD | Controller | Virtual Research V8 HMD | Optitrack | Near Far |
| 5 | [30] | Raj et al. | Desktop VR | Object | Computer | Kinect v2 InterSense inertia cube3 | Far |
| 6 | [31] | Ha et al. | AR HMD | Gesture | ACCUPIX my bud | PTAMM camera tracking | Near |
| 7 | [32] | Vuibert et al | Desktop VR | Gesture-Object | Vision RF shutter glass | Optitrack Flex: V100 Motion capture | Far |
| 8 | [33] | Wang et al. | VR HMD | Object-Touch | eMagin z800 HMD | PhaseSpace motion capture | Near |
| 9 | [34] | Feng et al. | Fish tank VR | Object-Touch | Nvidia 3D Vision glasses | Polhemus Fastrak | Near |
| 10 | [35] | Mendes et al. | VR HMD | Object-Touch | Gear VR, Samsung s6 | Kinect v2 | Near |
| 11 | [36] | Krichenbauer et al. | VST-HMD | Object | Oculus Rift, OVR vision stereo | Leap Motion | Near Far |
| 12 | [37] | Ro et al. | AR HMD | Gesture-Touch | HoloLens | Kinect v2 | Far |
| 13 | [38] | Kim et al. | VR HMD | Controller | Oculus Rift Consumer | Positional | Near |
| 14 | [39] | Schlunsen et al. | VR HMD | Controller | HTC Vive pro | Leap Motion | Far |
| 15 | [40] | Heinrich et al. | Projected AR | Object | Projector | Fusion track 500 | Near |
| 16 | [41] | Sun et al. | Web VR | Mouse | Computer | NA | Far |
| 17 | [42] | Andersen et al. | AR HMD | Gesture | HoloLens 1 | 6-DoF V | Far |
| 18 | [43] | Liu et al. | VR HMD | Gaze | HTC Vive Pro-Eye | 6Dof VR, eye tracking | Far |
| 19 | [44] | Weiß et al. | AR HMD/VR/ Projected AR | Object | HTC Vive, HoloLens, projector | Vuforia image target, | Near |
| 20 | [45] | Fuvattanasilp et al. | Handheld AR | Touch | Apple iPad Pro (2017) | AR marker | Far |
| 21 | [46] | Lee et al. | VR HMD | Controller | HTC Vive | 6-DoF VR | Near Far |
| 22 | [47] | Yu et al. | VR HMD | Gaze | Pico Neo 2 Eye | 6-DoF VR, eye tracking | Far |
| 23 | [48] | Dastan et al. | VR HMD | Controller | Oculus Quest 2 | 6-DoF VR | Near |
| 24 | [49] | Ganias et al. | VR HMD | Controller | HTC Vive Pro Eye | Two base stations | Near |

The literature on the design of TOTTA is very varied, and the impact of widgets and UI is crucial yet underestimated, especially considering that they can improve user performance, usability, comfort, and ease of use [50, 51, 52]. Poor design can also have side effects, like distraction from the primary task and increased mental and physical demands [53, 54, 55].

In the past, 3D tracking technology was the main TOTTA limiting factor due to low precision, accuracy, and latency [56], [57]. However, MR technology is evolving rapidly, and more precise and cost-effective devices are expected [58]. Even assuming ideal tracking, the basic knowledge, methods, and tools for TOTTA guidance interface design (UI) are still open research [59, 60]. TOTTA literature is scattered into different application domains, different research objectives (hardware validation, widget validation), and MR setups and lacks well-established guidelines.

To our knowledge, no previous literature review addressed TOTTA positioning widgets. The main contributions of this work are:
C1: Define the TOTTA problem and its importance,
C2: Provide a systematic review of TOTTA visual widgets,
C3: Classify and compare the approaches,
C4: Discuss open issues and trends,
C5: Guides for further research for optimal design TOTTA,
C6: Interest in MR developers in many domains to save time and resources and create effective, easy-to-use spatial interfaces and applications.

The research questions that motivate the literature research on the precise TOTTA spatial visual guidance in MR are: "What are the existing approaches in literature?", "How are the TOTTA widgets designed?" and "How are TOTTA widgets evaluated?".

## 2 METHODOLOGY SYSTEMATIC REVIEW

Being a multidisciplinary and cross-domain problem, the query definition is divided into four categories: "spatiality," "task," "technology," and "objective" (Fig. 2). We used the Preferred Reporting Items for Systematic Reviews and Meta-Analyses (PRISMA) guidelines for conducting a systematic review [61].

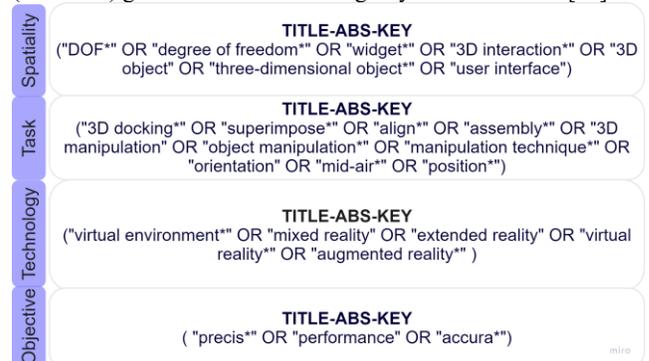

Fig. 2. The research query of the systematic review.

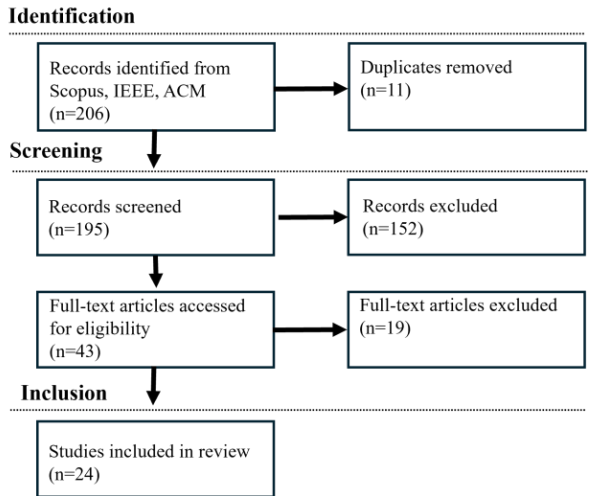

Fig. 3. PRISMA flow chart approved by [61].

This method is often used in conducting literature reviews [62, 63, 64]. We launched the search on the Scopus, ACM, and IEEE databases in August 2023. The PRISMA-based data collection stage is summarized in Fig. 3. From the initial 206 papers; we eliminated the papers that do not present or describe the widgets, the ones that focus on only hardware devices without user interfaces, and tasks that do not include a tool or target widgets. These aspects were essential in our systematic review and pointed out the key connection of papers, that they all present a spatial manipulation task of a visual real/virtual tool to a real/virtual target using TOTTA visual widgets. These criteria enabled detailed manual reading of the final selection of 24 papers in Table 1.

### 2.1 Bibliometric analysis

We visualized the keywords of the selected papers using word clusters [39] (Fig. 4). This graph shows that "3D user interfaces" and "3D interaction" are the key topics, and "computer-assisted surgery" and "object manipulation" are common applications. The colors demonstrate how recent studies (yellow) VR first, then AR, later in green to "user studies "and visualization, as the effect of natural technology evolution and optimization.

Fig. 5 supports the topic's novelty and demonstrates a positive trend. More than 50 percent of the found literature papers have been published in the last five years, and the increased number of citations and the majority of the papers are conference proceedings (16/24 papers, 67%).

The application domain is general in most of the cases (19/24 papers, 80%), with some verticalizations in medical (2) and industrial (3). This can be explained by the fact that TOTTA is a very common task, and these widgets can be useful in many domains.

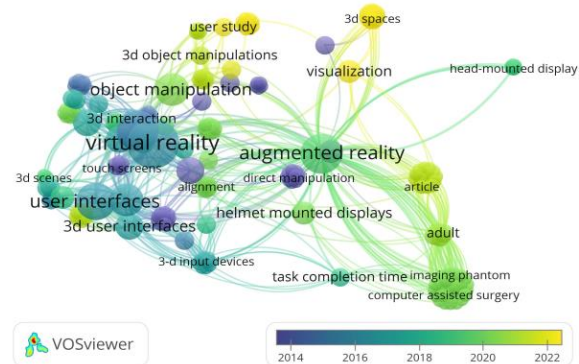

Fig. 4. Bibliographic keywords connection of the TOTTA papers.

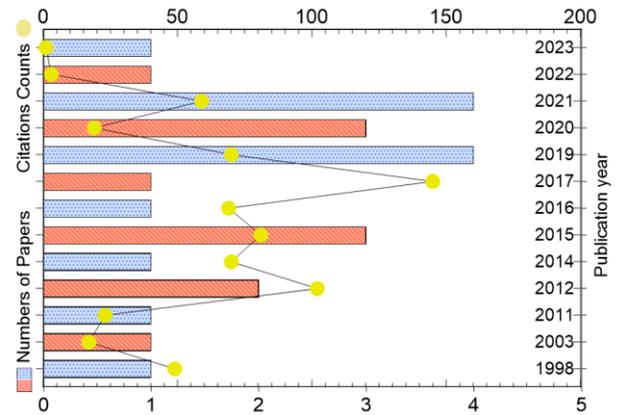

Fig. 5. The 24 TOTTA paper's publication year and citations.

### 2.2 TOTTA widgets in literature

This section presents the key aspects of the TOTTA widget design in chronological order, along with the main research drivers and findings.

Boritz et al. [26] propose a direct midair TOTTA interface using a new tetrahedron-shaped physical input device. The TOTTA widgets, which mimic the input device, are two identical tetrahedrons (1.73cm height and 0.87cm width), with one face perpendicular to the base and with a checkerboard texture (Fig. 6). The user must align TO on TA geometric shapes. When the distance is less than a threshold (0.5 cm), a red box appears at the tip of the TA. They found higher positional error along z (front direction of depth) with the monoscopic display vs stereo.

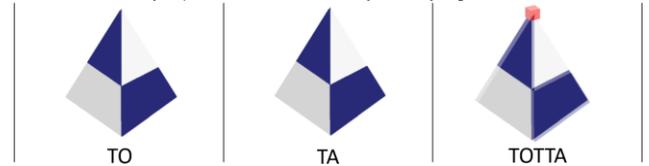

Fig. 6. Boritz et al. [26] use Tetrahedron, No: 1.

Fiorentino et al. [27] implement TO with a 3D Model - a simplified representation of the physical pen held by the user–; for TA, a green Box with colored 3D Axes (Fig. 7). With only a 3D positioning test, they find that difficulty varied with the target position: the targets in front of the user and above the head lead to greater error.

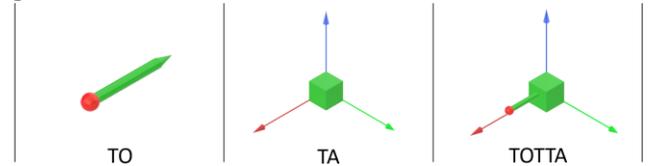

Fig. 7: Fiorentino et al. use Model TO, Box TA and Axis TA.

Veit et al. [28] investigate touch screen interactions with DOF separation (height axis vs. depth axis) using a monochromatic (green) 3D Axes TO (15 cm length) and a semi-transparent blue globe TA (7.5 cm radius). At collimation, the TO 3D Axes flash (Fig. 8). Results indicate that isolating the depth axis manipulation increases precision, while haptic cues do not improve user precision.

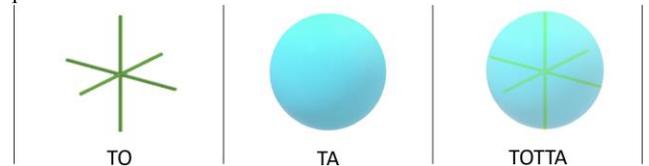

Fig. 8. Veit et al. [28] use 3D Axes TO and Globe TA, No:2.

Ragan et al. [29] propose a multi-touch input device for TOTTA. They use two Boxes of different sizes: TO is opaque blue, and the TA is semi-transparent purple, with eight small gray spheres in the vertexes. As additional feedback, the spheres turned red when only a vertex was aligned and green when all eight were aligned (Fig. 9). They found that touch-based interfaces increased the task completion time compared to wand or joystick interactions.

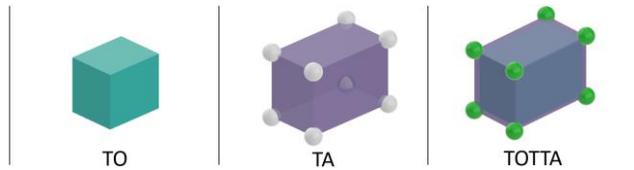
Fig. 9. Ragan et al. [29] use Box No: 3.

Raj et al. [30] utilize two different colored 3D Axes: TO, purple, and TA, gray. Both 3DAxes had unique-colored spheres at the endpoints to match the orientation (Fig. 10). The results differed between participants' gender and video game experience. This paper shows how the user's gender, avatar representation, and experience can influence performance. The self-avatar visualization resulted in a slightly faster rotation time than a sphere visualization of an avatar.

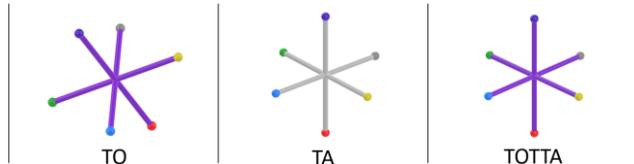
Fig. 10. Raj et al. [30] use 3D Axes, No:4.

Ha et al. [31] implemented a bare-hand user interface with a green Box for TO and a blue Box for TA combined with semi-transparent grey guidelines and shadows where the lines intersect in the AR environment (Fig. 11). The TA Box turns red as the target achieved feedback. This design is interesting as it claims that anteroposterior depth visual feedback by shadows and guidelines enables precise manipulation.

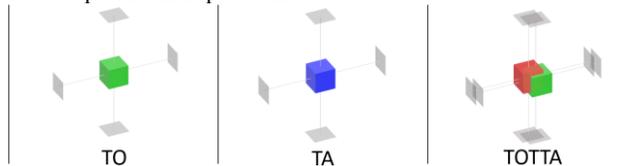
Fig. 11. Ha et al. [31] use Box No: 5.

Vuibert et al. [32] compared the tetrahedrons versus two chair models in desktop VR; they use a pair of same-sized Tetrahedrons rendered as a colored wireframe (Fig. 12). TO has an opaque sphere in the center, and TA has a larger semi-transparent sphere. This research found that virtual 3D models can perform better (time and precision) than tetrahedrons, probably by leveraging natural human skills from the real world. However, the 3D Model must allow unique positional and angular collimation.

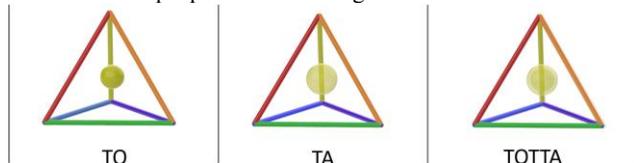
Fig. 12. Vuibert et al. [32] use Tetrahedron, No: 6.

Wang et al. [33] developed an Object Impersonation metaphor that enables switching the DRIVE (avatar view on a tablet and tetrahedron's view on HMD) and VIEW methods (avatar view on HMD and tetrahedron's view on a tablet). They use a pair of same-size Tetrahedrons with small colored spheres in the vertexes and additional Crosshair (Fig. 13). This widget comprises two coplanar, perpendicular lines (forming a 2D cross) surrounded by concentric circles. TO is semi-transparent blue with gray edges and orange crosshair, and TA is non-transparent turquoise with non-transparent crosshair. The object impersonation method gave better orientational precision but required higher cognitive demand.

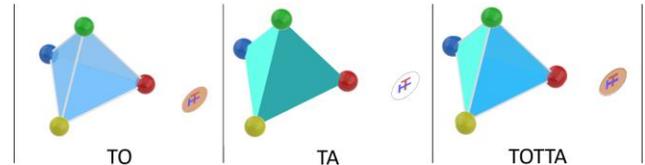
Fig. 13. Wang et al. [33], Tetrahedron & Crosshair, No: 7.

Feng et al. [34] evaluate one-hand free vs novel two-handed input devices for 7 DOF manipulation techniques. The input device corresponds to virtual spherical cursors (blue-left, pink-right). They use Box TOTTA with colored faces, but different in size and frame color: red (TO) and white (TA) (Fig. 14). The virtual cursors are linked with an orange-colored cylinder "spindle" with a small red sphere mid-point. The technique's result is equivalent when the TOTTA size is the same and faster with their input device when the TOTTA size is different.

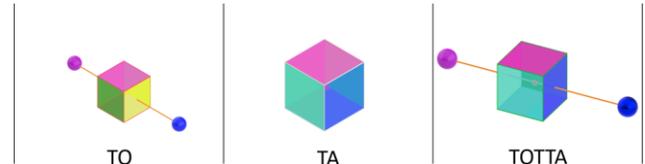
Fig. 14. Feng et al. [34] use Box, No:8.

Mendes et al. [35] present an opaque 3D Model with colored 3D Axes with sphere endpoints for TO, and a transparent 3D Model for TA (Fig. 15). Interactive secondary feedback is provided by the Model TO color gradually turning green with the distance from TA. An interesting aspect is the 3D axes allowed to control a single DOF. The PRISM technique dynamically adjusts between hand and object motion ratio. Experimentation demonstrated how DOF separation brings benefits but at the cost of task completion time.

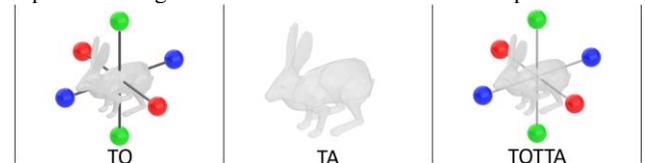
Fig. 15. Mendes et al. [35], 3D Axes, No: 9.

Krichenbauer et al. [36] compare AR vs VR, TO as a composition of an opaque Box with a single red face textured with the number two, colored 3D Axes, and a wireframe-colored Globe. (Fig. 16). TA is a box like TO with different sizes, transparency, and dashed edges. The results claim that AR resulted in faster completion time than VR when using a 3D input device and mouse.

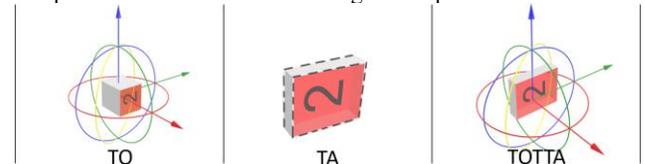
Fig. 16. Krichenbauer et al. [36], 3D Axes/Globe/Box; No:10.

Ro et al. [37] present a novel physical input device using the Laser pointer metaphor AR pointer. The user collimates the Box TO (green, blue, red) remotely with a touch on the mobile device to a different size semitransparent TA (Fig. 17). The AR pointer performs better in task completion time than the direct free hand 3D manipulation metaphor.

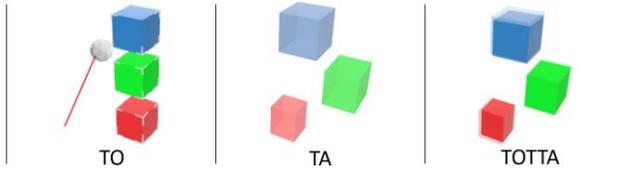
Fig. 17. Ro et al. [37] use Box No: 11.

Kim et al. [38] compared "DOF *separation* (1DOF, only axis-handles)", "*without* DOF *separation* (3DOF, only center-handle)," and "switchable DOF (1-DOF/2-DOF/3DOF, center, and axis-handles)" for mid-air manipulation. They utilize TO as a combination of a 3D Model (a teapot), colored 3D Planes, and colored 3D Axes with spheres in the endpoint (Fig. 18). TA is represented only as a semi-transparent 3D model. During manipulation, the constrained axis becomes yellow along lines or planes. The switchable DOF outperformed others in terms of time and precision efficiency.

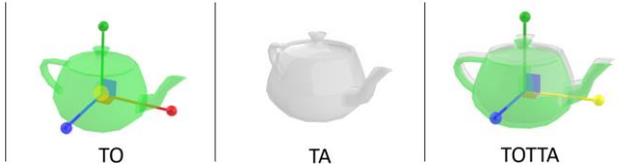
Fig. 18. Kim et al. [38] use 3D Axes/Plane/Model, No: 12.

Schlunsen et al. [39] evaluate free hand vs widget-based manipulation techniques and different multimodal cues for 3D manipulation of system control tasks. Green framed gray Box TO with 3D Axes TO (translation) with spheres in the endpoints and gray framed Globe TO (rotation) is used (Fig. 19). TA is a brown framed yellow box. The free-hand manipulation resulted in faster and most preferred by the participants. They claim that multimodal feedback (audio) improved the user experience.

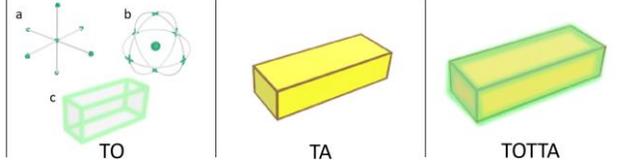
Fig. 19. Schlunsen et al. [39], 3D Axes/Globe/Box, No: 13.

Heinrich et al. [40] compare three visual widgets (circle, the crosshair, and the arrow concept) for AR-supported medical needle insertion (Fig. 20). They use different-sized 2D Crosshair TOTTA, small crosshair TO (color change red-orange-green), and bigger crosshair (transparent- yellow-green- red for depth feedback). Each concept has distinct color mapping and indicator scaling. The Crosshair outperformed in orientation and depth parameters. The results for the color and indicator scaling factors are less consistent.

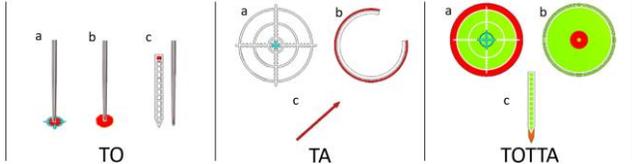
Fig. 20. Heinrich et al. Center TOTTA, No: 14.

Sun et al. [41] compare DOF manipulation modes in WebVR to explore user workload and task performance effects. TO use colored 3D Axes (translation) with cones in the endpoints and Globe TO (rotation) with three wireframes with different colored rings with spheres of interaction points (Fig. 21). The manipulated axis appeared yellow while others disappeared (1DOF). Multiple DOFs provide less perceived workload and higher presence. The results indicated that users feel less workload or more presence and tend to spend less time completing tasks on WebVR.

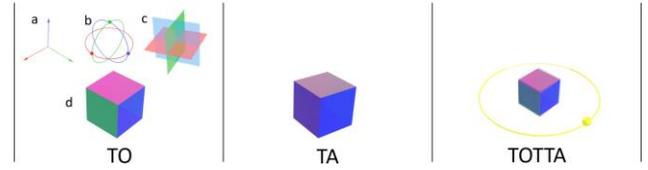
Fig. 21. Sun et al. [41] use Sphere/Axes/Plane, No: 15.

Andersen et al. [42] elaborate on three semi-transparent widget designs for mid-air interaction. The 3D Axes TOTTA, Crosshair TOTTA (white and red), and triangular pyramid TOTTA (Fig. 22). The crosshair and triangular TOTTA have the shortest alignment time. In contrast, 3D Axes TOTTA performs best in translation and rotation errors. A novel aspect is that visual elements' size affects how far the user extends the arm, influencing torque forces.

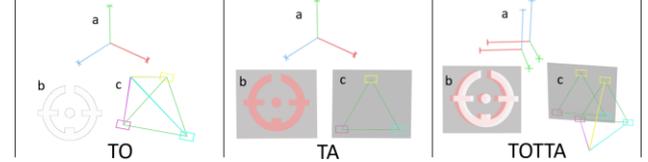
Fig. 22. Andersen et al. [42] use Crosshair, No: 16.

Liu et al. [43] evaluate which type of gaze-based manipulations (eye vs. head) performs best when combined with OrthoGaze. OrthoGaze allows the user to manipulate gray Box TO using the orthogonal Planes TO. The TA is represented as the green-colored Box TA (Fig. 23). During TOTTA, the gaze-selected Plane activates, and the user adjusts the 2-DoF position by looking at the target location and confirming placement through a gaze dwell. The eye gaze results are more accurate than the head gaze for continuous aiming.

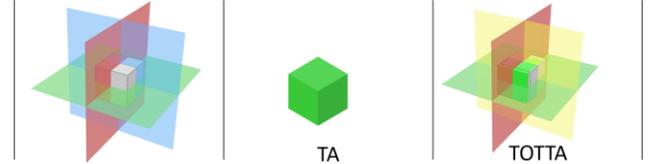
Fig. 23. Liu et al. [43] use Box and Plane, No: 17.

Weib et al. [44] investigate DIY tasks such as woodworking (drill, saw, and screw) using various levels of guidance: 2D video instructions, VR, and AR. They use distinct-size Crosshairs for TO and TA (Fig. 24). The TOTTA gives dynamic feedback to the user to avoid user error during drilling. The results indicate that context-aware situated visualizations are less likely to rely on empirical methods.

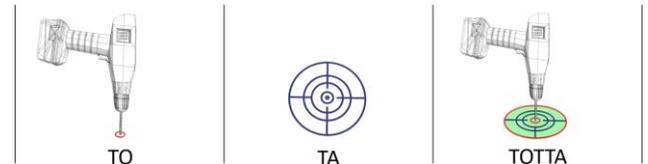
Fig. 24. Weib et al. [44] use Crosshair, No: 18.

Fuvattanasilp et al. [45] implemented SlidAR+, a novel handheld AR device (HAR) with an interaction method. They use a red-colored Line TO and a translucent, green-colored arrow with a thin blue line TA (Fig. 25). The user matches the TO (red arrow) with the TA (green arrow with the base of the virtual pillar). To align perfectly, a red line appears from the arrow TO tip to the TA, and the user slides TO using the Line. SlidAR+ resulted in faster task completion and is preferred by users.

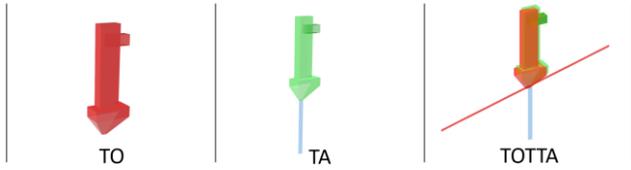
Fig. 25. Fuvattanasilp et al. [45] use Line, No: 19.

Lee et al. [46] propose a novel near-field interaction metaphor for distant object manipulations. They compare widget-based metaphor with unimanual metaphor (one hand & scaled replicated model) and bimanual metaphor (both hand & scaled replicated model). The Box TO has eight pair-colored small spheres in the vertex points (Fig. 26). For translation, a colored 3D Axes TO is used with a small yellow cube visually indicating the manipulated axis and red text for numerical feedback. For rotation, a colored wireframe Globe TO with a sphere is used. TA is a white wireframe with colored spheres on vertex points. The unimanual metaphor has the highest efficiency, the widget-based metaphor has the slowest, and the bimanual metaphor, with a scaled replica, grants the lower movement. Interestingly, subjective impressions are most favorable with the bimanual metaphor.

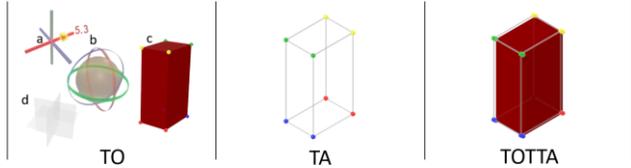
Fig. 26. Lee et al. Sphere/3D Axes/Plane TOTTA, No: 20.

Yu et al. [47] compare four gaze-supported interaction techniques: gaze grab, remote hand, 3D Magic gaze, and implicit gaze. They used a 3D Model for TO (a rabbit) and a transparent replica (Fig. 27). During the manipulation, the TO has a blue outline, and when the target is achieved, it turns red. The results indicate that gaze does not influence performance when the TO is in front of the user, but it can be useful for distant targets and larger spaces. The gaze input reduces the arms fatigue issue and potentially allows future TOTTA manipulation.

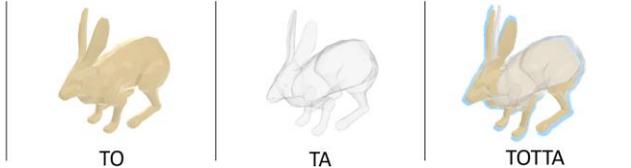
Fig. 27. Yu et al. Model TOTTA, No:21.

Dastan et al. [48] present a 5DOF guidance applied to dental implantology as the rotation along the drill axis is not influent. TO comprises three triangle pairs - colored differently for each direction- and two semi-circle pairs - colored differently for each rotation (Fig. 28). The pairs visualize in real time and amplify the position and rotation distance values. TA is a static green line with a concentrical cylinder. This approach leverages human reification from Gestalt theory [34], seeking a quick, pre-attentive reaction. Their method performed better in angular (with major effects) and positional precision and accuracy, with less mental demand and frustration than the literature. However, this gain is obtained with a significant increase in task time and physical demand.

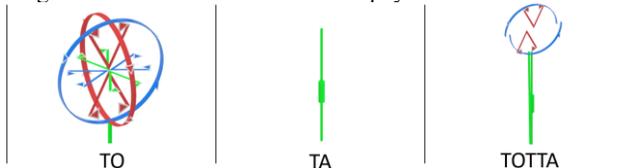
Fig. 28. Dastan et al. Axes/Globe/Line TOTTA, No:22.

Ganias et al. [49] compare grasping visualizations, auto-pose (realistic grasp), single pose (hands do not penetrate with object), and disappearing hand (hand disappears). They used a colored Box for TO (solid) and a transparent replica for TA (Fig. 29). At the end of each positioning, the visual cue of the changed color was used (the yellow returned green). The results indicated no significant difference in user performance in any visualizations. The auto pose is a user preference and provides a stronger perceived sense.

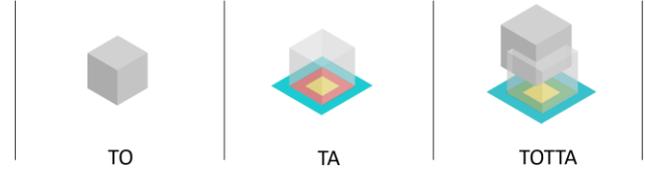
Fig. 29. Ganias et al. Box TOTTA, No:24.

## 3 Widget design analysis

A key aspect of TOTTA widgets is the distance of interactions. Near-field interactions allow the ability to direct manipulation of tool and target in proximity and are advantageous for the precision of small-size objects. On the other hand, the far-field interactions rely on distant object manipulations beyond the user's arm reach, which is advantageous for large tools and target size and distance flexibility. There is no prevalence between the near-field (12/24 papers, 50%) or far-field interaction (9/24 papers, 37%), and few studies (3/24 papers, 13%) employ both (Tab. 1).

The TOTTA may require various levels of DOF, so we analyze that it requires superimposing or aiming. The most common widget guidance mechanic is the visual 3D superimposition of TO over TA (20/24 papers, 83%) (Fig. 30).

A limited number also includes the scaling (8/24 papers, 33%) with the superimposing task, which consists of changing the size of the TO widget to match the TA. Visual scaling has no direct meaning for TOTTA collimation, even if it may have the potential to set some tool parameters (e.g., drill speed value). However, this usage is not envisioned in the selected papers.

Few papers prefer the aiming task (4/24 papers, 17%), which provides a reference point for the user to aim and align the TO used for 5DOF tasks (e.g., needle insertion, drill positioning).

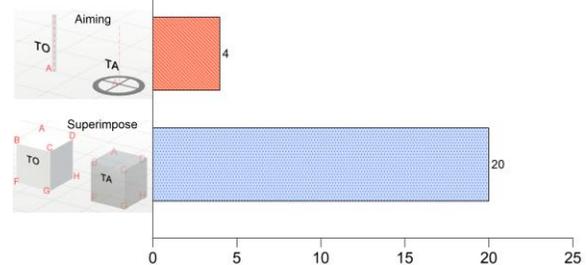
Fig. 30. Three main TOTTA tasks in the evaluations.

For a precise TOTTA, users receive multimodal guidance feedback associated with TOTTA widgets. At first glance, TOTTA visual widgets use basic shapes. However, each TOTTA representation is unique in detail. Therefore, the whole design gathered in a TOTTA widget cluster, resulting in 70 different graphical designs (Fig. 1).

In the next subsections, we analyzed the widget guidance mechanics through their feedback to the user. We divided the TOTTA widget design as feedback factors, "Collimation Feedback," "End of task feedback," and lastly, "During manipulation feedback."

### 3.1 Collimation Feedback

The TO and TA shapes must provide positional and orientation graphical/geometrical clues for unique collimation configuration. This aspect is not trivial and is key in widget design. We found eight diverse types of shapes: Box, 3D Axes, 3D Model, Crosshair, Globe, Tetrahedron, Lines, and Planes (Fig. 31).

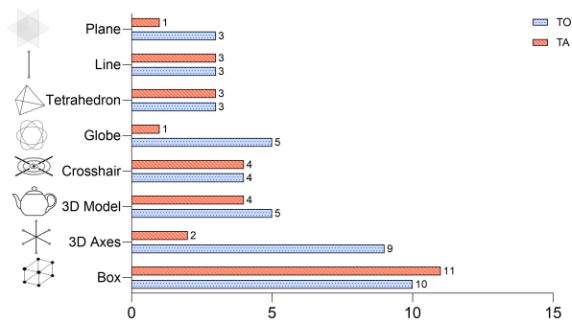

Fig. 31. Eight different visual designs were used for the TO and TA widgets.

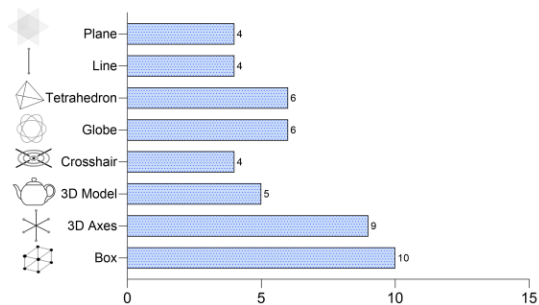

Fig. 32. Single basic shape type used by literature.

The box is the most preferred geometry (21/70 design, 30%, 10 TO/11 TA). It is simple, has predictable orthogonal angles, and is easy to implement in all graphical engines. The Box design is preferred by 10/24 papers [28, 30, 33, 36, 37, 38, 40, 42, 45, 49] (Fig. 32). It's also curious how boxes are used in physical child toys in learning motor skills. Box widgets with different dimensions can provide unique positional reference and partial angular. For angular, unique primary Box is supported by additional shapes or different colored/textured faces, edge styles, or sizes.

3D Axes are also common (11/70 design, 16%, 9TO/2TA), with a clear and familiar design for CAD users and gamers. The 3D Axis design is preferred by 9/24 papers [27, 34, 35, 38, 40, 41, 45, 48, 65] (Fig. 32). The presence of a center supports positional placement, and the orthogonal lines facilitate angular arrangements. TO and TA are differentiated by colors or additional geometries, like spheres in the vertices.

3D Models (9/70 design, 13%, 5TO/4TA) (e.g., rabbit, teapot, arrow) instead of basic geometries. This design is preferred by 5/24 papers [34, 36, 37, 44, 46] (Fig. 32). TO and TA 3D Models are commonly differentiated by color and transparency. However, the choice of a specific model is limited as some 3D models may be inefficient in positional and orientational guidance. Therefore, they are substandard or rarely used by the end user.

Crosshair (8/70 design, 11%, 4TO/4TA) is common in aviation, military, and healthcare interfaces. The Crosshair design was chosen from 4/24 papers [32, 39, 41, 43] (Fig. 32). Planar or 3D crosshairs are used in spatial interactions with reduced DOF, like needle insertion and drilling.

Globe (6/70 design, 9%, 5TO/1TA) is a familiar widget design for desktop applications and games. Interestingly, 6/24 papers (Fig. 32) preferred Globe and they are often preferred singularly only TO or only TA without having them together in the task [27, 35, 38, 40, 45, 48]. Globe is often represented as wireframe rings or semi-transparent to avoid visual occlusion.

The tetrahedron (6/70 design, 9%, 3TO/3TA) is geometrically the simplest ( having minimum entities) shape for the TOTTA widget. 3/24 papers use it [21, 27, 28] (Fig. 32). The corners help with orientation but are less familiar compared to the other shapes and have no orthogonal angles.

Line (6/70 design, 9%, 4TO/2TA) ) can provide single DOF guidance with reference to mid/end points by distinctive styles (dashed, solid, transparent, or color). However requires the combination of elements to be functional in the 3D space. The Line widget is preferred by 4/24 papers [33, 36, 41, 46] (Fig. 32).

Planes (4/70 design, 6%, 2TO/1TA) are usually represented in three perpendicular surfaces. Their intersection can also generate 3D axes. However, Planes are less intuitive and more prone to visual occlusion than others. The Planes is preferred by 4/24 papers [38, 41, 43, 46] (Fig. 32). Some other visual representation methods are used in TOTTA to support collimation feedback (Fig. 33).

As theorized by perception, the same or similar shapes for TO and TA are used principally (22/24 papers, 92%).

To further distinguish the tools from the targets, colored minor parts of widgets (such as frames/vertices/faces) (18/24 papers, 75%) or entirely colored widgets (17/24 papers, 71%) are used frequently.

Transparency (16/24 paper, 67%) is also used to reduce visual clutter and enhance depth perception during collimation.

Following, TO\TA pairs are differentiated by containing additional/different geometrical elements like small spheres and cubes (14/24 papers, 59%).

Some preferred (13/24 papers, 54%) more than one design of widgets, we defined them as "Mixed widgets", [38, 39, 46]. They have used the composition of shapes (from Fig 31), requiring more cues for guidance—even redundant in some cases—since they can result in more complexity.

Finally, the different widget sizes (11/24 papers, 46%) are used to distinguish TO from TA, especially if the task requires resizing.

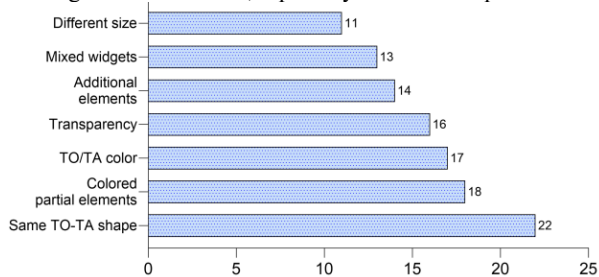

Fig. 33. Collimation feedback mechanics: graphical clues.

### 3.2 During Manipulation Feedback

This feedback supports the user during the TOTTA task through continuous guidance feedback (Fig. 34). Only 14/24 literature papers use during manipulation continuous feedback; they are interactive and not just signals. The majority of these TOTTA designs use continuous color change (e.g., red-orange-green) to convey guidance (9/24 papers, 13%).

Text feedback (2/24 papers, 13%) is also preferred to indicate the real-time error values. Some use sonification (2/24 papers, 8%), which is a simultaneously generated sound (e.g., drums with variable rhythm) to guide the user interactively [32, 39]. Text feedback can provide precision control during TOTTA. However, since there is a continuous change of text in the field of view, it may frustrate the user or cause a high task load.

As dynamic novel approach visual error visualization (1/24 papers, 4%) is used during the manipulation. This approach promises intuitive feedback; it allows dynamic visualization of the target distance by widgets' forms of distance [48]. Another unique aspect is that the widgets disappear at the target threshold, reducing clutter. The widgets' more dynamic and complex behavior is demonstrated to improve user performance but at the cost of physical and cognitive demand. This aspect of widget design has an undisclosed potential to guide the user along the interaction in addition to collimation.

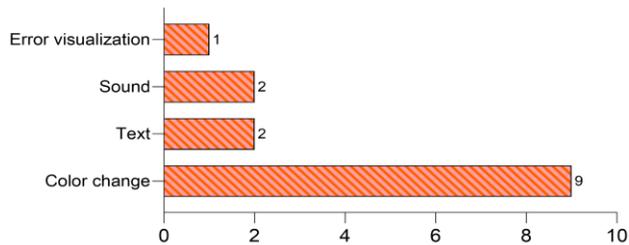
Fig. 34. During manipulation, continuous feedback.

### 3.3 TA Reached Feedback

As the TO and TA shapes begin to superimpose, visual collimation feedback becomes increasingly inefficient. Therefore, "supplementary" feedback is often provided concurrently with the collimation (Fig. 35). This feedback conveys the end-of-task information as signals and they are triggered at the TOTTA positional and rotational distance threshold, probably to substitute haptic feedback (e.g. vibration) [31, 38, 39, 43].

Commonly instant color change is used after the task completion (17/24 papers, 71%) as it is intuitive for the user and is easy to implement. Further, at the collimation event, playing a completion sounds (9/24 papers, 38%) followed by a new object/target appearance (6/24 papers, 25%) [33] and text flash (1/24 papers, 4%) "Right There!".

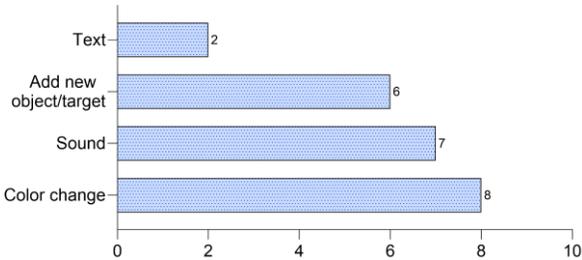
Fig. 35. TA reached Feedback at the end of the task.

## 4 EVALUATION METHODS

TOTTA widgets perform differently and can be physically and mentally demanding for the user. Not much research has been done to highlight how the literature compares and evaluates different TOTTA designs. Therefore, in this section, we analyzed the literature evaluation methods in four subsections: research questions, Participants, Procedure, and Metrics.

### 4.1 Research Questions

Common literature research questions are manipulation methods (10/24 papers, 42%): DOF separation, multi-level DOF, learning (knowledge retention, skill acquisition, and transferability), or user experience evaluation (presence and engagement), Fig. 36.

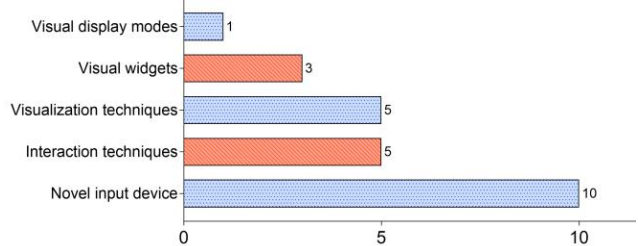
Fig. 36. Research Questions of the TOTTA papers.

A secondary research question concerns interaction devices (5/24 papers, 21%), such as hand-held controllers, finger-tracking, and eye-tracking, and their effect on user experience regarding usability, user satisfaction, and performance. A third common research question concerns visualization techniques (5/24 papers, 21%) and the effects of AR/VR user experience on presence and engagement.

Our research discovers that the research questions on widget visual design are limited (3/24 papers, 13%) despite its significant impact on the user experience and performance.

Other research questions are specific: gender video game experience affects self-avatar representation (1/24 papers, 4%) on the 3D manipulation task.

### 4.1 Participants

Within-subject (20/24 papers, 83%) is the most utilized methodology among the literature papers. The average number of participants is 21 +—SD 12.9, with a minimum age of 18 and a maximum age of 50. Most are right-handed (92%), young (SD 3.1, mean 25.8), and male (65%) (Fig. 37). The majority are student participants (90%), some unpaid volunteers (4 /24 papers), and some paid (4/ 24 papers), while the rest is not specified.

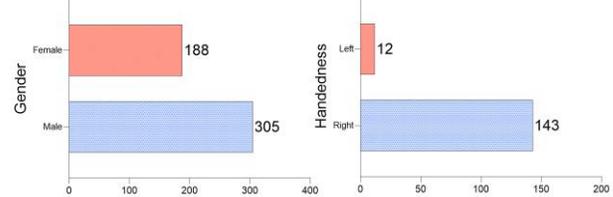
Fig. 37. Participants' demographic information.

### 4.2 Procedure

TOTTA validations are more commonly tested in VR (19/24 papers, 79%) than in AR. Twelve papers use VR HMD, five desktop VR, three AR HMD, two Projected AR, one Fish Tank VR, one VST-HMD, and one hand-held AR, as shown in Table 1.

Often the participants are provided with a hand-held device (17/24 papers, 71%) since it simulates the feeling of a TO being picked up or provides physical feedback when interacting with an input device or controller, Table 1. A custom-made object input device is the most preferred (9/24 papers, %37), followed by controller (7/29 papers, %29) and mouse (2/24 papers, %8) input devices. Moreover, 5/24 papers (%21) involved a free-hand interaction, and 2/24 papers (%8) included a gaze interaction as well. In some cases, the participants were also provided with virtual avatars and input device representation (8/24 papers, 33%), which influences self-perception and the sense of presence in mixed realities.

### 4.3 Metrics

The literature papers measure performance, acceptance, and preference using similar metrics. The most inquired metric is task completion time (24/24 papers, 100%), followed by subjective data (20/24 papers, 83%), positional error (15/24 papers, 63%), angular error (11/24 papers, 46%), scale error (4/24 papers, 79%), and count (e.g., attempts, click) (4/24 papers, 17%) (Fig. 38).

We analyzed the Subjective data since it is essential for comprehensively understanding the user experience, needs, emotions, and behaviors collected during the experiment. The most investigated metric is user preference (14/24 papers, 58%), followed by the perceived performance (9/24 papers, 38%), NASA-TLX (7/24 papers, 29%), difficulty rate (6/24 papers, 25%), comfort rate (6/24 papers, 25%), SUS (4/24 papers, 17%), intuitiveness (4/24 papers, 17%), PQ-presence questionnaire (3/24, 13%), UEQ (2/24 papers, Fig. 39. The least used metrics are AttrakDiff (1/24 papers, 5%), ARI (1/24 papers, 4%), and HARUS (1/24 papers, 4%).

## 5 DISCUSSION

This systematic review showed that TOTTA widget designs are non-standard and differ in design, feedback, and interaction. In all studies, the TOTTA guidance is supported by the Gestalt theory's visual overlap or cognitive psychology as a "collimation feedback," intuitive and pre-attentive feedback.

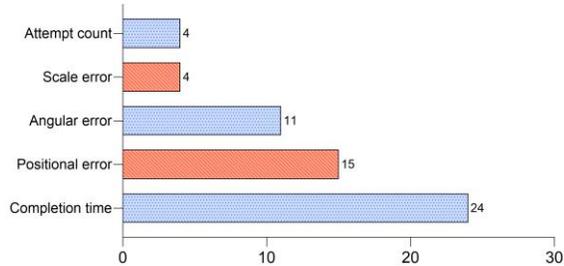

Fig. 38: Quantitative TOTTA collected metrics.

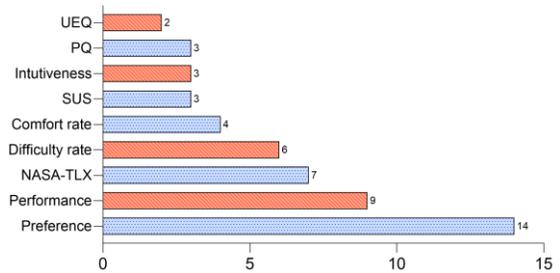

Fig. 39. The subjective TOTTA collected metrics.

The visual overlap of similar basic shapes is supported by reification, a pre-attentive human capability of interpreting visual information as theorized by the Gestalt laws, such as proximity, closure, similarity, and continuation.

This aspect is not always supported from a theoretical\ cognitive point of view in the studies. Box, 3D Axis, and 3D Models are the most preferred ones. However the geometries that are missing a center (e.g., wireframe Box) can have problems in precise positioning.

The alignment of the same basic-shaped TO and TA presents the challenge of differentiating them. Some evaluated approaches are frames/vertices/faces color or style (75%), different-colored TO/TA (71%), transparency (67%), additional shapes (59%), multiple widgets (54%), and size (46%). Using colored, partial elements and transparency, TO and TA are differentiated. Transparency can also be beneficial for reducing visual occlusion during the TOTTA task.

Although some patterns are visible and probably lead to effective solutions, few studies provide scientific ground for visual and interaction design. We argue that a deeper comprehension of the perception of shapes (e.g., Gestalt theories) can lead to better results in terms of performance and usability.

Another aspect is the common design of the continuous "during manipulation feedback." Sonification and animations can bring large margins of improvement. The latest papers are evolving from the one-time signal at goal reach to a potentially more effective, dynamic, and responsive guidance method.

Another key finding is the lack of a well-established golden standard and direct comparisons of the present widgets or a partial set. The hardware systems used in the selected papers vary in terms of immersion, tracking, and visual quality, deeply impacting the resulting experience, user performances, and acceptance.

The TOTTA validation methods found in this systematic review use quite different -thus not comparable-experiment designs (target configuration, avatar, background, etc.), and we spot some bias (VR is more tested than AR, right-handed, and male participants). Another aspect to highlight is that the accessibility issues are not investigated or mitigated (e.g., color code and color blindness), and none of the studies reviewed addressed this topic.

### 5.1 Limitations and Future Works

This study acknowledges its limitations because no previous research has been done in the literature to analyze the TOTTA widget design and evaluation methods. We examined papers that required a specific tool to align targets using visual widgets. Tasks such as assembly were excluded since the user searches for tools and targets in the environment and constructs a whole. This limitation allows us to analyze each method deeply.

For future works, it is important to evaluate the possible trade-off between the widget complexity and the cognitive overload as experienced in some experiments [33, 48]. We think that to achieve better performance, as requested by the industry, widget behavior will increase in complexity, and the future challenge is to balance this with users' cognitive overload.

Also, DOF separation improved user precision so that it may be investigated better as a design solution in future studies. On the grounds of this study, we can draw the future TOTTA research:
- Define and implement a standard experiment framework.
- Equally compare the existing TOTTA widgets.
- Improve widget design by perception theory.
- Improve widgets with continuous guidance.
- Enforce diversity in user tests.

### 5.2 Key findings

We extracted some main research outcomes and key findings approved by TOTTA papers, considering that they are valid in the specific context, widget design, and experimental conditions.
- TOTTA stereoscopic performs better than monoscopic view [26].
- 6DOF direct hand manipulations [29, 39], DOF separation [28, 35], switchable DOF [38], increased precision, completion time, and qualitative results.
- Self-avatar visualizations reduce task time [30], and participants preferred the semi-transparent hand [31].
- Object impersonation (user embodiment in TO) provides better orientational error but increases cognitive demand than DRIVE and VIEW metaphors [33].
- If TO vs. TA sizes differ (the user must also apply to scale), the physical input device influences the task precision and time, and bimanual interactions are better suited than unimanual [34].
- The participants perform faster in AR than in VR with a 3D input device [36].
- Free-hand manipulation is faster than widget-based manipulation and is preferred by the participants, and multimodal cues improve the user experience [39].
- The 2D crosshair performs better than the 2D arrow for translation and rotation [40].
- 3D Model TOTTA performs better than 3D Axes/Globe [46].
- The tetrahedron shape has better orientational precision, but the task time increases [32] compared to the model shape.
- Gaze-based manipulation causes more fatigue than controller-based manipulation [43].

## 6 CONCLUSION

This work defined the TOTTA guidance problem and its importance in several applications. We provided a systematic review, starting from 206 papers, and we selected 24 papers that deeply analyzed dedicated widgets. We described the existing approaches, studied TOTTA widget design characteristics and evaluation methods, and presented key findings of the selected papers. We are encouraged by this study's findings, which identify the gaps and provide context for future research. This research provides interesting perspectives and guides for the many research applications and industries where the precise tool to target object manipulation is required, such as medicine, aviation, industry, retail, etc. This study concludes with acknowledgment of the need to address TOTTA problems and the necessity of a standardized but flexible evaluation methodology due to the different declinations of MR systems in the scientific community.

ACKNOWLEDGMENTS

Co-founded by MICS (Made in Italy - Circular and Sustainable) - National Recovery and Resilience Plan (PNRR), 4-2-1.3 - No. 341, Italian Ministry of University and Research, European Union – NextGenerationEU PE00000004, CUP D93C22000920001. Co-funding by Emme Evolution S.r.l. FESR 2014 – 2020 Obiettivo Convergenza. Project development organizations: with Polytechnic University of Bari, CUP B95H22000810007.